\documentclass[aps,prx,10pt,showpacs,amsmath,showkeys,floatfix,amssymb, preprintnumbers, nofootinbib, superscriptaddress,longbibliography]{revtex4-1}
\usepackage{times}
\usepackage{amsmath}
\usepackage{amssymb}
\usepackage{epsfig}
\usepackage{color}
\usepackage{makecell}
\usepackage{subcaption}

\begin{document}

\title{Scattering matrix elements and energy spectrum of one-dimensional hybrid $\mathcal{P}\mathcal{T}$-symmetric finite systems}
\author{Vladimir~Gasparian}
\email{vgasparyan@csub.edu}
\affiliation{Department of Physics and Engineering,  California State University, Bakersfield, CA 93311, USA}

\author{Esther~J\'odar}
\email{esther.jferrandez@upct.es}
\affiliation{Departamento de F\'isica Aplicada,  Universidad Polit\'ecnica de Cartagena, E-30202 Murcia, Spain}

\author{Antonio Pérez-Garrido}
\email{antonio.perez@upct.es}
\affiliation{Departamento de F\'isica Aplicada,  Universidad Polit\'ecnica de Cartagena, E-30202 Murcia, Spain}

\begin{abstract}
In this work, we provide a complete description of the scattering matrix elements and electron energy spectrum in one dimensional $\mathcal{PT}$-symmetric hybrid finite systems, using the characteristic determinant approach. We present an analytical formulation of the problem and obtain a closed-form expression for the energy spectrum of the system, consisting of a region of real potential (passive region) surrounded by regions of gain and loss on the left and right, respectively. It has been shown that under certain conditions and a specific ratio between the real and imaginary parts of the complex potentials, it is possible to find analytical expressions for the spectral singularities at which the scattering matrix elements of the hybrid structure tend to infinity at a specific
real energy. Within the framework of the same approach, we present a compact analytical expression for the quantization condition that determines the energy spectrum of a model corresponding to the placement of a rigid lattice within a finite-sized box.
\end{abstract}

\pacs{42.25Fx,42.70Qs,42.79Ek}

\maketitle

\section{Introduction}
Non-Hermitian Hamiltonians describe the
phenomenological behavior of experimental systems involved in scattering or decay processes.
Such systems receive energy from and/or deposit energy into their environment, so they are
typically not in equilibrium, the total energy and probability are not conserved, and the energy
levels are complex numbers; the imaginary part of an energy level is associated with the lifetime
of the physical state or the width of a scattering resonance.

However, in $\mathcal{P}\mathcal{T}$-symmetric systems the gain from the environment and loss to the environment
is exactly balanced. As a consequence, even though they are not isolated, $\mathcal{P}\mathcal{T}$-symmetric systems
behave like Hermitian systems in that they are in equilibrium and their energy levels are real (so called $\mathcal{P}\mathcal{T}$ symmetric phase). In this case the phase of the system can be regarded as the $\mathcal{P}\mathcal{T}$ symmetric phase. However, the $\mathcal{P}\mathcal{T}$ symmetry broken phase is characterized by complex eigenvalues (see e.g., Refs. \cite{real,real1}).

In this work, we have provided a complete description of the scattering matrix elements and electron energy spectrum in one
dimensional  $\mathcal{P}\mathcal{T}$-symmetric hybrid finite systems, using the characteristic determinant approach. 
While the fact that the scattering matrix elements and the energy spectrum of electrons are closely related to each other seems natural from the point of view of the Green's function approach (both are the poles of the Green's function for the open and closed systems, respectively), it is a highly non-trivial task from the point of view of technical calculations: one needs quite substantial analytical and numerical efforts in order to correctly obtain the critical behavior of the spectrum quantitatively.

In Section III (Energy Spectrum), when studying the energy spectrum of the hybrid system, we will see that idealized closed systems, described exclusively by the Schr{\"o}dinger equation, and open systems, described by non-Hermitian Hamiltonians, differ from each other only in the initial conditions of the characteristic determinant, or, in terms of the $2\times 2$ transfer matrix method, in the initial conditions of the element $M_{22}$. The determinant approach, in principle,  is compatible with the transfer matrix method and is convenient for both numerical and analytical
calculations, including very large systems, where the boundary and initial conditions
are not as important. 

The method of the characteristic determinant used is, in principle, capable of providing an answer to a large class of questions related to the quantum transport of particles (waves) and Ising model with an arbitrary nearest-neighbor random exchange integral, temperature, and random magnetic field in one or two-dimensional systems (disordered or not). 
This approach allows expressing the transmission coefficient $T_N(k)=|D_N|^{-2}$ of a wave propagating through an $N$-layer system or a set of $N$ delta potentials in terms of the determinant $D_N$, which depends only on the reflection amplitudes from an individual scatterer. The zeros of the characteristic determinant (which coincide with the zeros of the inverse transmission amplitude ${1}/{t_N(k)}$ or, in terms of the $2\times 2$ transfer matrix method, with the zeros of the element $M_{22}$) coincide with the poles of the Green's function and play an important role in calculating the energy spectrum of a closed system (see references \cite{Gas2025,AGG91}). Knowledge of the explicit form of determinant $D_N$ allows us to study the density of states averaged over the sample as well as the energy spectrum of excitations in the layered structure and also their propagation (surface polaritons and plasmons, etc)\cite{AGG91}. All 
these results
are obtained by expressing the characteristic determinant as a Toeplitz tridiagonal or five-diagonal Toeplitz determinant.

The paper is organized as follows. In Sec. II the model $\mathcal{P}\mathcal{T}$-symmetric hybrid finite systems and some useful relations concerning the characteristic determinant
approach and the scattering matrix elements are introduced and discussed. We will assume that the strengths of two Dirac functions $V_1$ and $V_2$, characterizing the gain and loss regions on the left and right, are complex of the form $V_1 =\eta_1 + i \eta_2$
and $V_N=V_1^{*}=\eta_1-i\eta_2$, respectively. The passive region describes with $m$ real $V_l$ delta potentials with coordinate $x_l$, which are symmetrically distributed around the midpoint. We obtained closed-form expressions for the singularities of the scattering matrix elements in the general case when $\eta_1 \ne 0$ in Sec. II (C)  
In Sec. III we investigate the energy spectrum of the hybrid system by imposing boundary conditions in the form of hard walls at points $x_0$ and $x_{N+1}$, adding two extra delta potentials $V_0$ and $V_{N+1}$. 
This is followed by the discussions and summary in Sec. IV.

\section{Model}

\subsection{Real potential region (passive region) surrounded by gain and loss regions on the left and right}

The model has been the subject of numerous numerical studies, similar in spirit but differing in detail (see, for example, references \cite{japan1,2,xx} and references therein). However, a general analytical description of the energy spectrum as a function of system-characterizing parameters such as gain, losses, and passive regions remains lacking.
The aim of this paper is to address this problem and attempt to find analytical results for two closely related quantities, such as the scattering matrix elements (open system) and the spectrum for a one-dimensional delta potential model (see Eq. (\ref{pot2})) in a closed system . To our knowledge, such calculations have not been reported previously. To this end, we extend the characteristic determinant-based approach
to discuss and adequately describe the spectral properties of the finite hybrid $\mathcal{PT}$-symmetric system. Before doing so,
we briefly recall some instructive analytical results obtained for the characteristic determinant $D_N$ and present the derivation of the spectrum of a hybrid structure consisting of a real potential region (passive region) surrounded by gain and loss regions on the left and right, respectively.

A scattering potential $V(x)$, satisfying $\mathcal{P}\mathcal{T}$-symmetric conditions, that we will use to describe the scattering matrix elements $t$, $r_l$, $r_R$ and as well as the energy spectrum of the hybrid system, can be represented as
a sequence of $N$ $\delta$-potentials  $V_l$, 
\begin{equation}
V(x)=V_1\delta(x-x_1) + V_N\delta(x-x_N)+
\sum_{l=2}^{N-2}V_l\delta(x-x_l), \label{pot2}
\end{equation}
where the first $V_1$ (for $x = x_1$) and the last $V_N$ (for $x = x_N $) delta potentials are complex of the form $V_1 =\eta_1 + i \eta_2$
and $V_N=V_1^{*}=\eta_1-i\eta_2$, respectively. The passive region (third term) is described with $N-2$ real $V_l$ delta potentials with coordinate $x_l$, which are symmetrically distributed about the midpoint.

The most efﬁcient way to calculate the reflection and transmission amplitudes, $r_L$, $r_R$, $t$ and the energy spectrum
for the given potential, in our point of view, is the characteristic determinant $D_N$, which, as was mentioned above, is closely related to the transfer matrix method and introduced in Ref.\cite{GA88} ($k=\sqrt{
E}$ in units $2m_0 = 1$ and $\hbar = 1$):
\begin{equation}
D_N=\det\left| \delta_{nl}+\frac{iV_l}{2k}\exp{ik|x_l-x_n|}\right|. \label{e:det}
\end{equation}
The characteristic determinant, $D_N$ can be written as the 
determinant of a tridiagonal Toeplitz matrix and satisfies the following
recurrence relationship:
\begin{equation}
D_N={A_N} D_{N-1} - {B_N} D_{N-2}\;,\label{e:d0}
\end{equation}
where $D_{N-1}$ ($D_{N-2}$) is the determinant Eq.(\ref{e:d0})
with the $N$th (and also the $(N-1)$--th) row and column omitted.

The coefficients $A_N$, $B_N$ can be obtained from the explicit
form of $D_N$. For $N>1$ we have :
\begin{equation}
{A_N}=1+ {V_N\over V_{N-1}}e^{2ik(x_{N}-x_{N-1})}+{iV_N\over 2k}
\left[1-e^{2ik(x_{N}-x_{N-1})}\right],
\end{equation}
and
\begin{equation}
{B_N}={V_N\over V_{N-1}}e^{2ik(x_{N}-x_{N-1})}.
\end{equation}
The initial
conditions for the recurrence relations are:
\begin{equation}
D_0=1,\qquad D_{-1}=0,\qquad {A_1}=1+{iV_1\over 2k}.
\end{equation}

The transmission amplitude $t$ is the inverse of the
characteristic determinant $D_N$ multiplied by the phase accumulated
during the transmission,
i.e.,
\begin{equation}
t={e^{ik(x_N-x_1)}D_N^{-1}}, \label{t}
\end{equation}

As for the reflection amplitude $r_L$ for electrons incident from the left is obtained from the following relationship:
\begin{equation}
r_L={2k\over iV_{1}}{D_N-D_{-1+N}\over D_{N}}-1 \equiv
{i}{\partial \ln t \over \partial \frac{V_1}{2k}}-1, \label{rL1}
\end{equation}
where $D_{-1+N}$ is the characteristic determinant without the first delta
function (i.e. $V_1=0$).
Similarly, the reflection amplitude $r_R$ for the incident wave from the right reads 
\begin{equation}
r_R={2k\over iV_{N}}{D_N-D_{N-1}\over D_{N}}-1 \equiv
{i}{\partial \ln t \over \partial \frac{V_N}{2k}}-1, \label{rr1}
\end{equation}
where $D_{N-1}$ is the characteristic determinant without the last delta
function (i.e. $V_N=0$).

Our aim now is to write explicitly the dependence of the characteristic
determinant $D_N$ on
${V_1}$ and $V_N$ for our general system of $N$ $\delta$-potentials (see Eq. (\ref{pot2})). Based on Refs. \cite{gas1,GAM} and 
using the above recurrence relations for the
characteristic determinant, applied to both ends, we can
rewrite $D_N$ in the following way ($m=N-2$) (see Refs. \cite{GAM,Gas2025})
\begin{equation}
D_N(V_1,V_N)=D_{m}f_m(V_1,V_N)
\label{dn}
\end{equation}
where $f_m(V_1,V_N)\equiv 1+A_mi\frac{V_1}{2k}+B_mi\frac{V_N}{2k}+C_m{\frac{V_1}{2k}}{\frac{V_N}{2k}}$ and
$D_{m}$ is the characteristic determinant for the previous
potential without the first and the last delta function (i.e.,
$V_1=V_N=0$). $A_m$, $B_m$ and $C_m$ are coefficients
independent of $V_1$ and $V_N$ and involving $D_{m}$ (for details see Ref. \cite{Gas2025} and references therein).
These coefficients are defined as

\begin{align}
    \left.
        \begin{aligned}
            & A_m= 1-i\sqrt{1-T_m}e^{i\Theta_1}\\
            & B_m= 1-i\sqrt{1-T_m}e^{i\Theta_2}\\		
            & C_m =2ie^{i\frac{\Theta_1+\Theta_2}{2}}\bigg[\sin{\frac{\Theta_1+\Theta_2}{2}}
+\sqrt{1-T_m}\cos{\frac{\Theta_1-\Theta_2}{2}}\bigg]\\
				\end{aligned} 
				\right\}. \label{Aa}  
\end{align}

Here $T_m=t_mt^{*}_m$ is the transmission coefficient of the block with $m$
delta-potentials. The phases appearing in the previous equations are defined as
$\Theta_1=\varphi_m+\varphi_{am}+2k(x_2-x_1)$ and
$\Theta_2=\varphi_m-\varphi_{am}+2k(x_N-x_{N-1})$. $\varphi_m$ is the phase accumulated in a
transmission event and $\varphi_{am}$ is the phase
characterizing the asymmetry between the reflection to the left
and to the right from the block with $m$ delta-potentials.

The total transmission coefficient $T_N$, using Eq. (\ref{t}), 
can be represented in a quite compact form:
\begin{equation}
T_N=\frac{1}{|D_N|^2}=\frac{T_m}{f_mf^{*}_m}, \label{tx1}
\end{equation}
where $T_m=\frac{1}{|D_m|^2}$ is the transmission coefficient through the system with $m$ delta potentials without the first and last potentials. The expression (\ref{tx1}) of $T_N$ is exact, depends on  the sample geometry and valid for any value of the control parameters, $k$, $V_1$ and $V_N$  for a fixed $m$
and will be studied in detail below (see subsections $B$ and $C$).

Similarly, one can find the reflection amplitudes from the left $r_L$
and right $r_R$ incident waves. For example, $r_L$, using Eq. (\ref{rL1}), can be written as 

\begin{equation}
\begin{aligned}
r_L=-\frac{1+B_m{i}\frac{V_N}{2k}+(A_m-iC_m\frac{V_N}{2k})(i\frac{V_1}{2k}-1)}{1+A_mi\frac{V_1}{2k}+B_m{i\frac{V_N}{2k}}+C_m{\frac{V_1}{2k}}{\frac{V_N}{2k}}}
, \label{r_L2}
\end{aligned}
\end{equation}
where $A_m$, $B_m$ and $C_m$ are defined by Eq. (\ref{Aa}).

The amplitude of reflection $r_R$ from right can be calculated either directly using Eq. (\ref{rr1}), or replacing $V_N\leftrightarrow V_1$ and $B_m\leftrightarrow A_m$

\begin{equation}
\begin{aligned}
r_R=-\frac{1+A_m{i}\frac{V_1}{2k}+
(B_m-iC_m\frac{V_1}{2k})(i\frac{V_N}{2k}-1)}
{1+A_mi\frac{V_1}{2k}+B_mi\frac{V_N}{2k}+C_m\frac{V_1}{2k}\frac{V_N}{2k}}, \label{r_L1}
\end{aligned}
\end{equation}
It is easy to see by comparing the equations (\ref{r_L2}) and (\ref{r_L1}) that the reflection amplitudes falling from the left and from the right are not equal to each other $r_L\ne r_R$. This indicates a non-reciprocity of the reflection amplitudes. This is opposite to transmission amplitude $t_N$, defined by Eq. (\ref{t}), which is reciprocal.

One can check directly, using Eqs. (\ref{tx1}), (\ref{r_L2}) and (\ref{r_L1}), that the pseudounitary conservation
relations hold (see, e.g. Refs. \cite{xx,xy,zz,xz}):
\begin{equation}
|T_{N}-1|=\sqrt{R_{L}R_{R}}\label{ener} ,
\end{equation}
or 
\begin{equation}
T_{N}+r_{L}r^{*}_{R}=1,  \label{ener1}
\end{equation}
where $R_{L/R}\equiv |r_{L/R}|^{2}$ are the reflection coefficients from the left and right and $T_{N}=|t_{N}|^2$ 
defined above.

After a short introduction of some useful relations concerning the characteristic determinant approach and the scattering matrix elements, in the following subsections, we will discuss the main properties of the transmission $T_N$ for several general cases with arbitrary potential distributions, when the analytic expressions
for $T_m$ can be calculated explicitly.
\subsection{Periodic case}
First, let us consider a model of a passive region in which $m$ delta-function
potentials of an equal amplitude $V_0$ are located
periodically at $x = na$ ($a$ is the period of
the region): Let us assume also that $x_{m+1}-x_m=x_2-x_1=a$ and $\varphi_{ma}=0$ (that is, the scattering potential of the passive region is located symmetrically with respect to the first and last potentials). 
For this particular case of symmetry, the parameters defined in the Eq. (\ref{Aa}) can be written in a slightly simplified form ($\Theta_1=\Theta_2$):
$A_m=B_m=1-i\sqrt{1-T_m}e^{i\Theta}$, $C_m=2ie^{i{\Theta}}\bigg[\sin{{\Theta}}+\sqrt{1-T_m}\bigg]$ 
and
\begin{equation}
\Theta=\varphi_m+2ka =ka-\arctan \bigg[\bigg(\frac{V_0}{2k}\cos{ka}-{\sin {ka}}\bigg)\frac{\tan{(m\beta a)}}{\sin{(\beta a)}}\bigg],\label{theta}
\end{equation}
Recalling that $V_1=\eta_1+i\eta_2$ and $V_N=V^{*}_1=\eta_1-i\eta_2$, the characteristic
determinant $D_N(\eta_1,\eta_2)$ (see Eq. (\ref{dn})), becomes $D^{0}_mf^{0}_m(\eta_1,\eta_2)$ with 
$D^{0}_m$ and $f^{0}_m$, defined by 
\begin{equation}
D^{0}_{m}=e^{imka} \left \{\cos {(m\beta a)} + i\bigg(\frac{V_0}{2k}\cos{ka}-{\sin {ka}}\bigg){\frac{\sin{(m\beta a)}}{\sin{(\beta a)}}} \right\},
\label{Mn}
\end{equation}
\begin{multline}
f^{0}_m(\eta_1,\eta_2)=\frac{e^{i\Theta}}{2k^2}\bigg\{
2k^2\cos{\Theta}+2k
\bigg(\sin\Theta+\sqrt{1-T_m}\bigg)\eta_1\\
+i\bigg[2k\eta_1\cos{{\Theta}}-2k^2\sin{\Theta}+\bigg({\eta_1^2+\eta^2_2}\bigg)
\bigg(\sin\Theta+\sqrt{1-T_m}\bigg)\bigg]\bigg\}
\label{dn1}
\end{multline}
where 
\begin{equation}
\begin{aligned}
\cos ({\beta a})=\cos ka +{\frac{V_0}{2k}}\sin {ka}. \label{spectra}
\end{aligned}
\end{equation}
Note that in the forbidden gap $\cos{\beta a}$ must be replaced by $\cosh{\beta a}$.

By using the relation (\ref{tx1}) and the explicit expression of $D^{0}_m$ (\ref{Mn}), the total transmission coefficient through the system $T_N$ can be written in the form:
\begin{equation}
T_N=\frac{T_m}{|f^{0}_m(\eta_1,\eta_2)|^2}=\frac{1}{|f^{0}_m(\eta_1,\eta_2)|^2}\frac{1}{1+(\frac{V_0}{2k})^2\left({\sin m\beta a\over \sin \beta a}\right)^{2}}. \label{ty}
\end{equation}
where $T_m$ is the transmission coefficient through the system with $m$ delta potentials without the first and last potentials:
\begin{equation}
T_m=\frac{1}{|D^{0}_m|^2}=\frac{1}{1+(\frac{V_0}{2k})^2\left({\sin m\beta a\over \sin \beta a}\right)^{2}}. \label{tz}
\end{equation}
Note that the total transmission coefficient $T_N$,  (Eq. \ref{ty}), through the system continuously exceeds unity, if $|f^{0}_m(\eta_1,\eta_2)|^2<T_m$. This occurs only for energy $E$ near the band edges (a more
detailed numerical investigation of the behavior of $T_N$ will be presented below (see Figs.\ref{TN} and \ref{Fig4}).

Before concluding this section let us write down $r_{L/R}$ for the discussed model when the passive region has $m$ delta-function
potentials of an equal amplitude $V_0$ are located
periodically at $x = na$
\begin{multline}
\begin{aligned}
r_{L/R}=
\frac{-2ik^2e^{i\Theta}}{f^{0}_m(\eta_1,\eta_2)}\times \bigg[\sqrt{1-T_m}+\frac{\eta_1\cos\Theta}{k}+
\bigg(\sin\Theta+\sqrt{1-T_m}\bigg)\bigg(\frac{\eta^2_1+\eta^2_2}{2k^2}\pm\frac{\eta_2}{k}\bigg)\bigg], \label{r_LR}
\end{aligned}
\end{multline}
The peak structure of the reflection coefficient $R_{L/R}$ from left/right, based on the analytic formula (\ref{r_LR}), will be analyzed in more detail in the following subsection.
\subsection{Spectral Singularities}
It is known that two types of singularities are present in scattering amplitudes: 1) a branch cut sitting along the positive real axis in complex $k$-plane that separates physical sheet (the first Riemann sheet) and unphysical sheet (the second Riemann sheet); 2) poles of transmission and reflection amplitudes. These poles are called spectral singularities of a non-Hermitian Hamiltonian when they show up on the real axis, which yields divergences of reflection and transmission coefficients of scattered states (see, e.g., Refs. \cite{ahm1,must,gas1}).  

In the general case, the poles contain three unknown real variables $\eta_1$, $\eta_2$, and $k$, while the number of independent transcendental equations is two. 
Sometimes, obtaining a closed-form solution requires manually adding a third equation: this applies to the $\mathcal{P}\mathcal{T}$-symmetric hybrid finite systems considered in this article, and this will be explicitly demonstrated in subsection 2 ($\eta_1\ne 0$).

\subsubsection{$\eta_1=0$}

It is useful to consider as a starting point $\eta_1=0$,
where analytical expressions for the spectral singularities energies can be found relatively easily. It is easy to see that the zeros of $f^{0}_m(\eta_1=0,\eta_2)$ or ${f^{0}}^*_m(\eta_1=0,\eta_2)$ with respect to the variable $\frac{\eta^2_2}{2k^2}$ correspond to the cases where the transmission coefficient $T_N$ ($R_{L/R}=|r_{L/R}|^2$) tends to infinity are given by:
\[
\frac{\eta^2_2}{2k^2}\bigg|_{1,2}=\frac{\sin\Theta\pm i\cos\Theta}{\sin\Theta+\sqrt{1-T_m}},
\]
To obtain the real value of $\frac{\eta^2_2}{2k^2}$, it is necessary to substitute $\Theta\equiv \Theta_{cr}=\frac{\pi(2l+1)}{2}$ ($l=0,1,2,3\hdots$), which leads to the following result:

\begin{equation}
\eta_{2cr}=\frac{\sqrt 2 k_{cr}}{\sqrt{1+(-1)^l\sqrt{1-T_m}}}, \label{eta2}
\end{equation}
where $k_{kr}$ is the positive solution of the transcendental equation (\ref{theta}) with $\Theta_{cr}=\frac{\pi(2l+1)}{2}$. For this critical positive energy $E_{cr}=k^2_{kr}$, the transmission $T_N$ and reflection coefficients $R_{L,R}=|r_{L,R}|^2$
defined by Eqs. (\ref{ty}) and (\ref{r_LR}), become
infinite for a special critical value of $\eta_{2cr}$, defined above. Note that in particular case of $m=0$ or free propagation in the passive region when 
$T_m=1$ and $\Theta=ka$, the condition (\ref{eta2}) reduces to the single real solution $\eta_2=\frac{\pi(2l+1)}{\sqrt{2}a}$ at $ka=\frac{\pi(2l+1)}{2}$, discussed in Refs. \cite{must,ahm1}.

\begin{figure}
\includegraphics[width=8.4cm]{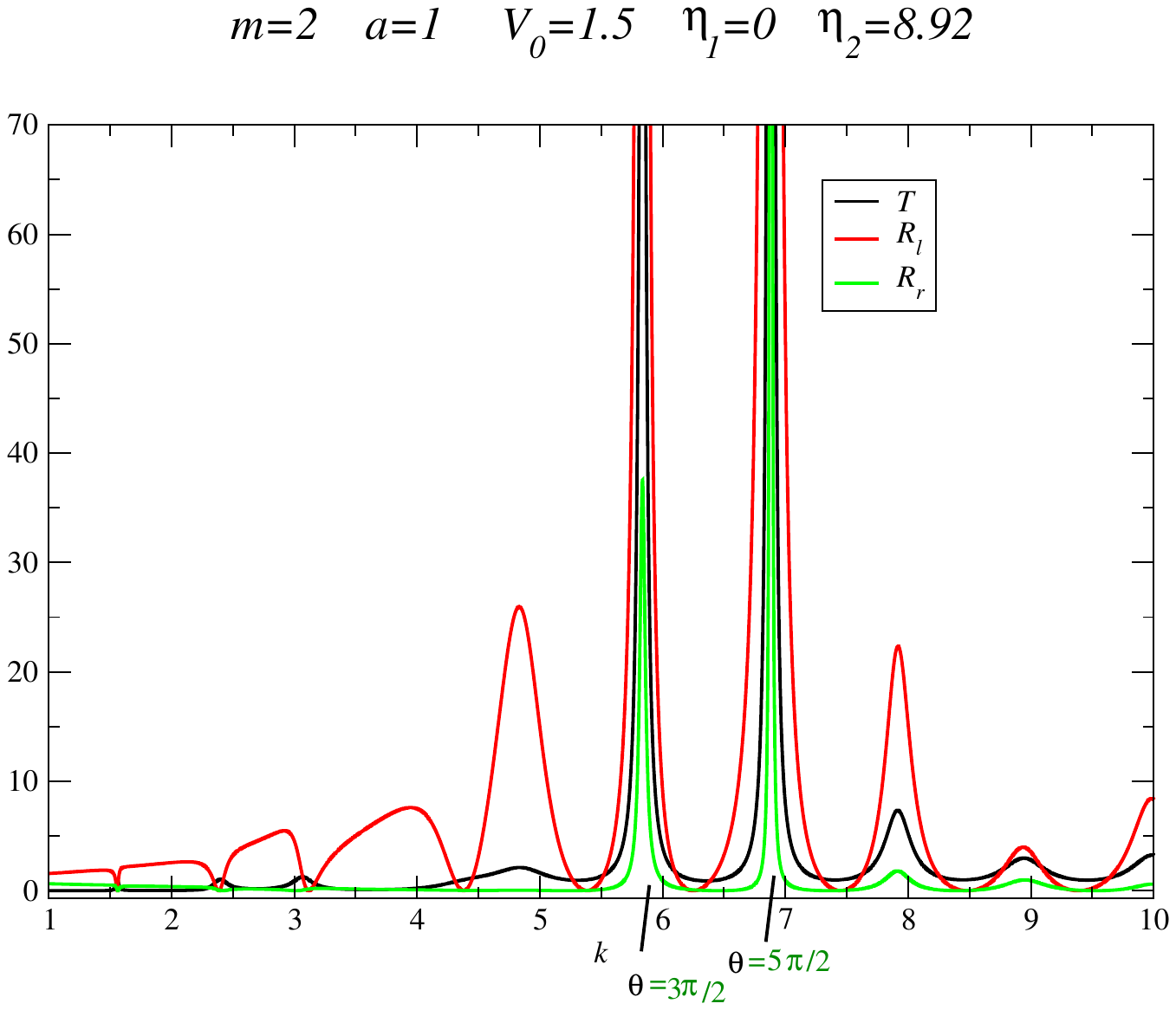}
\caption{The plot of $T_{N}$ and $R_{L,R}=|r_{L,R}|^2$ versus $k$ for the complex
$\mathcal{P}\mathcal{T}$ -invariant hybrid system. 
The passive region contains $m=2$ delta potentials. Two critical values of $\Theta$ at which divergence occurs are explicitly shown.} 
\label{Fig1}
\end{figure}

Fig. \ref{Fig1} shows the singularities of $T_N$ 
and $R_{L/R}$ in case of $\eta_1=0$,  confirming the theoretical prediction that the divergence
occurs for specific values of $\eta_2$ given in Eq. (\ref{eta2}) at the values
of $k_{cr}$ calculated exactly from Eq. (\ref{theta}). Two critical values of $\Theta_{cr}$ ($\frac{3\pi}{2}$ and $\frac{5\pi}{2}$) and two values of $\eta_2$ ($9.31$ and $8.92$, respectively) at which divergence occurs are explicitly shown. We note that a numerical analysis of expression (\ref{eta2}) shows that as $\eta_2$ decreases, for example $\eta_2=1.05$, the divergence may occur even at small $\Theta_{cr}=\frac{\pi}{2}$
($k_{cr}\approx 1$ not shown in Fig. \ref{Fig1}). 

\subsubsection{$\eta_1\ne 0$}
\begin{figure}
\includegraphics[width=8.4cm]{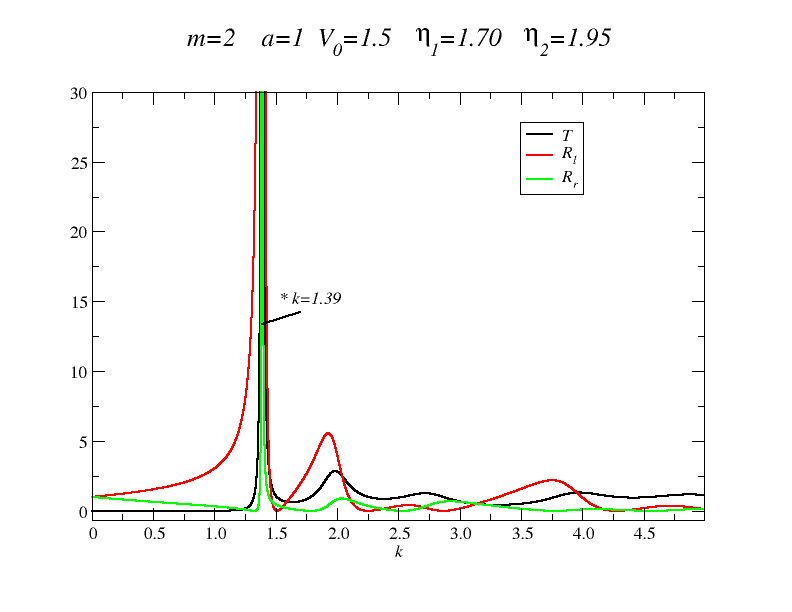}
\caption{The plot of $T_{N}$ and $|r(L/R)|^2$, versus $k$ for the complex
$\mathcal{P}\mathcal{T}$ -invariant hybrid system with $m=2$ delta potentials. The model parameters were chosen as follows: $V_0=1.5$ and $a = 1$. The calculation of the critical values $\eta_1=1.7$ and $\eta_2=1.95$ is based on Eqs. (\ref{kcr}) and (\ref{kcr1}).} 
\label{Fig2}
\end{figure}

The situation is somewhat more complicated in the case of $\eta_1\ne 0$, since the number of unknown real variables $\eta_1$, $\eta_2$, and $k$, as was mentioned, exceeds the number of independent transcendental equations. Interestingly, if we set 
\begin{equation}
2k^2\sin{\Theta}=\bigg(\eta_2^2-\eta^2_1\bigg)\bigg(\sin{\Theta}
+\sqrt{1-T_m}\bigg),\label{cond}
\end{equation}
then the hybrid model admits an exact solution, or more precisely, the denominator of (\ref{r_LR}) has $\bigg[k\cos {\Theta}+ \eta_1\bigg(\sin{\Theta}+\sqrt{1-T_m}\bigg)\bigg]$ as a factor. 
By solving the transcendental Eq. (\ref{cond}) for $k_{cr}$
\begin{equation}
\frac{k^2\sin{\Theta}}{\sin{\Theta}+\sqrt{1-T_m}}\bigg|_{cr}=\frac{\eta^2_2-\eta^2_1}{2},\label{kcr}
\end{equation}
and assuming that the factor
\begin{equation}
\bigg[k\cos {\Theta}+ \eta_1\bigg(\sin{\Theta}
+\sqrt{1-T_m}\bigg)\bigg]\bigg|_{cr}=0, \label{kcr1}
\end{equation}
simultaneously becomes zero at this point we 
find the pair of ($\eta_{1cr}$, $\eta_{2cr}$) and $k_{cr}$ at which the transmission and reflection coefficients, 
$T_N$ and $|r(L/R)|^2$,
become infinite. Fig.\ref{Fig2} confirms the divergent nature of the transmission and reflection  coefficients. At the critical point $k=1.39$, correctly predicted by Eq. (\ref{kcr}), they become infinite, providing an effective verification of the numerical results.

It should be noted that for the specific value $m=0$, i.e., $T_m\equiv 1$, the formula given above reduces to the result for two complex $\mathcal{P}\mathcal{T}$-symmetric Dirac delta potentials, obtained in Ref. \cite{ahm1} by directly solving the Schr{\"o}dinger equation.

After a brief discussion of the spectral features associated with the poles of the scattering matrix elements, we will then present a detailed and, in many respects, complete description of the behavior of the scattering matrix elements, ranging from the strong non-Hermitian regime to the weak non-Hermitian regime.
First, consider the case $\eta_1=0$ discussed in several papers \cite{japan1,2,xx}. It was shown in Ref \cite{2,aaa} that in a symmetric hopping tight-binding model with the amplitude of the jump between
two sites $t_h$, the transition from weak 
non-Hermiticity to the regime of strong non-Hermiticity is controlled by the ratio $\frac{\eta_2}{2t_h}$ ($\eta_2$ is a parameter that controls the gain/loss pair in the $\mathcal{P}\mathcal{T}$ system). The behavior of the transmission probability
is strongly non-Hermitian in the regime of weak non-Hermiticity with divergent peaks when 
$\frac{\eta_2}{2t_h}<1$ and is almost Hermitian in the regime of strong non-Hermiticity, $\frac{\eta_2}{2t_h}>1$ where the usual Fabry-Perot type peak structure is restored. 

For a simplified continuum model consisting of two complex delta potentials, it was found that divergent peaks of the transmission and reflection coefficients appear in the range $\frac{\eta_2}{2k}<1$ (see, e.g., Ref. \cite{2}). 

We now turn to a closer investigation of the various parameter ranges of the $\mathcal{P}\mathcal{T}$ -symmetric continuum model with passive region in the center. 
We will see that the parameter that will be responsible for the transition from Hermitian regime to the regime of strong non-Hermiticity is in general very complicated and depends on the number of potentials $m$ and other parameters, such $k$, $V_0$,  etc. The condition $\frac{\eta_2}{2k}<1$ (as well as $\frac{\eta_2}{2t_h}<1$)
arises as a limiting case of inequality (\ref{t13}) (see below) when
either the resonant case or free propagation in the passive region take place. For simplicity, let us consider the case of $\eta_1=0$.

\begin{figure}
\includegraphics[width=8.4cm]{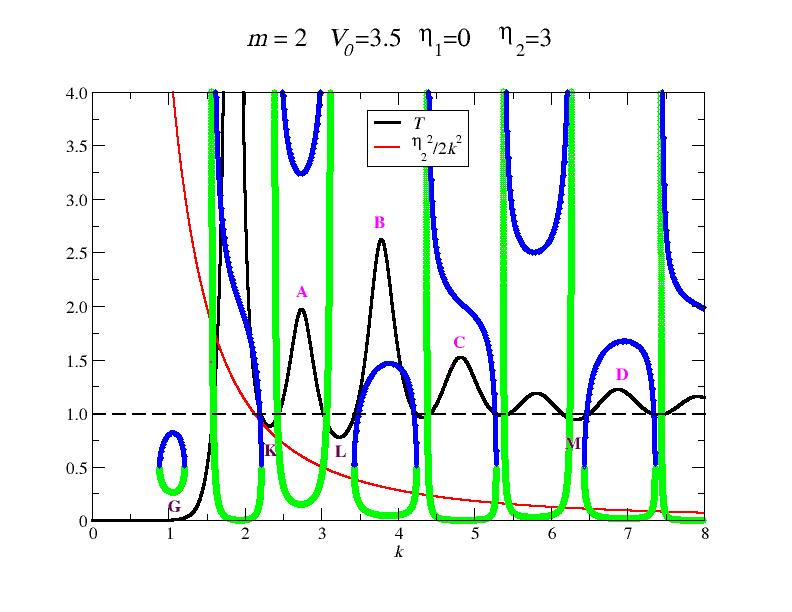}
\caption{The plot of $T_N$, Eq.(\ref{ty}), versus $k$ for the passive system with $m=2$.
The parameters are $\eta _1 = 0, \eta_2=3, V_0=3.5$, $a=1$. The red curve corresponds to $\frac{\eta^2_2}{2k^2}$, while the blue and green curves correspond to the right and left parts of the inequality, Eq. (\ref{t13}) respectively. Black curve is the total transmission coefficient $T_N$.}
\label{TN}
\end{figure}
\begin{figure}
\includegraphics[width=8.4cm]{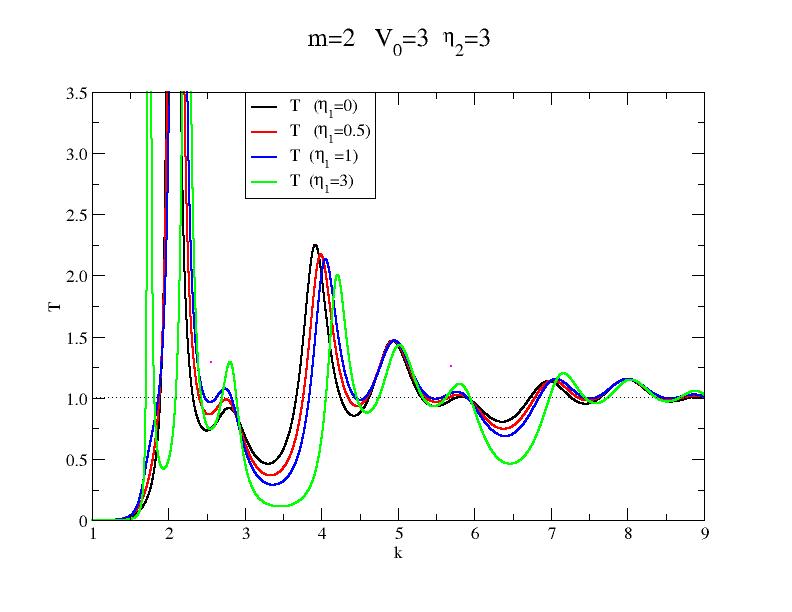}
\caption{The plot of Eq.(\ref{ty}) versus $k$ for the passive system with $m=2$ for different values of 
$\eta _1 = 0$ (black), $0.5$ (red), $1$ (blue), $3$ (green). The other parameters are: $m=2, \eta_2=V_0=3$, $a=1$. }
\label{Fig.5}
\end{figure}
As it was mentioned, it is clear that when  $|f^{0}_m(0,\eta_2)|^2<T_m$, the total transmission $T_N$ given by Eq.(\ref{ty}) is always greater than $1$. The appropriate ranges of $\frac{\eta^2_2}{k^2}$ defined as 
\begin{equation}
\frac{
{|\sin\Theta|-\sqrt{T_m-\cos^2\Theta}}}{|\sin\Theta|\pm \sqrt{1-T_m}}
<\frac{\eta^2_2}{2k^2}<\frac{{|\sin\Theta}|+{\sqrt{T_m-\cos^2\Theta}}}{|{\sin\Theta}|\pm\sqrt{1-T_m}}, \label{t13}
\end{equation}
provided the inequality $T_m-\cos^2\Theta\ge 0$ is satisfied. The $\pm$ sign corresponds to the ranges where $\sin\Theta \ge 0$ and $\sin\Theta < 0$, respectively. Fig.\ref{TN} displays the selected regions $A$, $B$, $C$ and $D$, where the total transmission coefficient $T_N>1$. As seen from Fig.\ref{TN}, $T_N$ shows one sharp peak (at $k\approx 2.2$) and several smooth and broad peaks of height larger than one with increasing $k$. For all these values of $k$, $|f^{0}_m(0,\eta_2)|^2$
goes through local minimums, starting from deep (small $k$) to shallow (large $k$). It is important to note that for regions A, B, C, and D, it is clearly visible that inequality (\ref{t13}) holds (Fig. \ref{TN}). The selected regions where the transmission coefficient is less than unity, $T_N<1$, and no longer show sharp peaks, are indicated by $G, K, L$ and $M$. In such cases, the conditions either $\frac{\eta^2_2}{2k^2}<\frac{|\sin\Theta|-\sqrt{T_m-\cos^2\Theta}}{|\sin\Theta|\pm \sqrt{1-T_m}}$ or $\frac{\eta^2_2}{2k^2}>\frac{{|\sin\Theta}|+{\sqrt{T_m-\cos^2\Theta}}}{|{\sin\Theta}|\pm\sqrt{1-T_m}}$ must be satisfied. In the regions $K, L,$ and $M$ shown in Fig.\ref{TN}, only the first inequality condition is satisfied, that is the red curve $\frac{\eta^2_2}{2k^2}$ is always below the green curve.
Unlike regions $K, L$ and $M$, in region $G$ the second inequality condition is satisfied, 
and, consequently, the red curve $\frac{\eta^2_2}{2k^2}$ is above the blue and green curves.
When $\frac{\eta^2_2}{2k^2}$ (red curve) crosses the green or blue lines in Fig. \ref{TN}, the total transmission coefficient $T_N$ becomes equal to one, that is, for these specific values of $k$, the coefficient $T_m$ through the passive region is equal to 
$|f^{0}_m(0,\eta_2)|^2$. For all the resonant values of $k$, where $T_N=1$ (black horizontal dashed line), $\frac{\eta^2_2}{2k^2}<2$, according to the inequality (\ref{t13}). At the special case when $T_m=1$ (either the resonant case or free propagation in the passive region when $m=0$), the above inequality takes a simple form: $0<\frac{\eta^2_2}{2k^2}<2$.
\begin{figure}
\includegraphics[width=8.4cm]{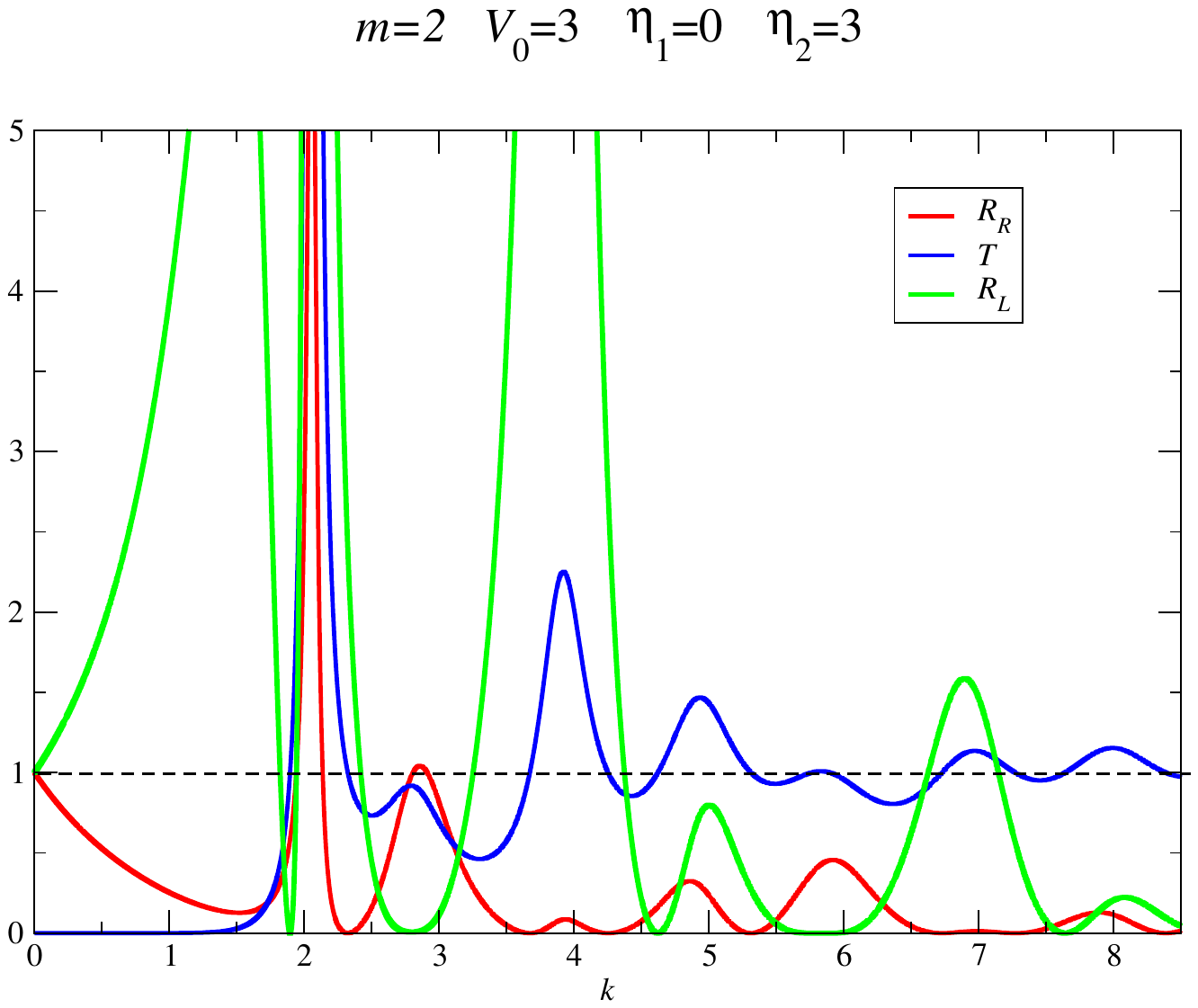}
\caption{The plot of $T_N$, Eq.(\ref{ty}), and $R_{L/R}$, Eq. (\ref{r_LR}), versus $k$ for the passive system with $m=2$ for $\eta _1 = 0$. The other parameters are: $\eta_2=V_0=3$, $a=1$. }
\label{Fig4}
\end{figure}

\begin{figure*}
 \begin{subfigure}[b]{0.99\textwidth}
\includegraphics[width=0.495\textwidth]{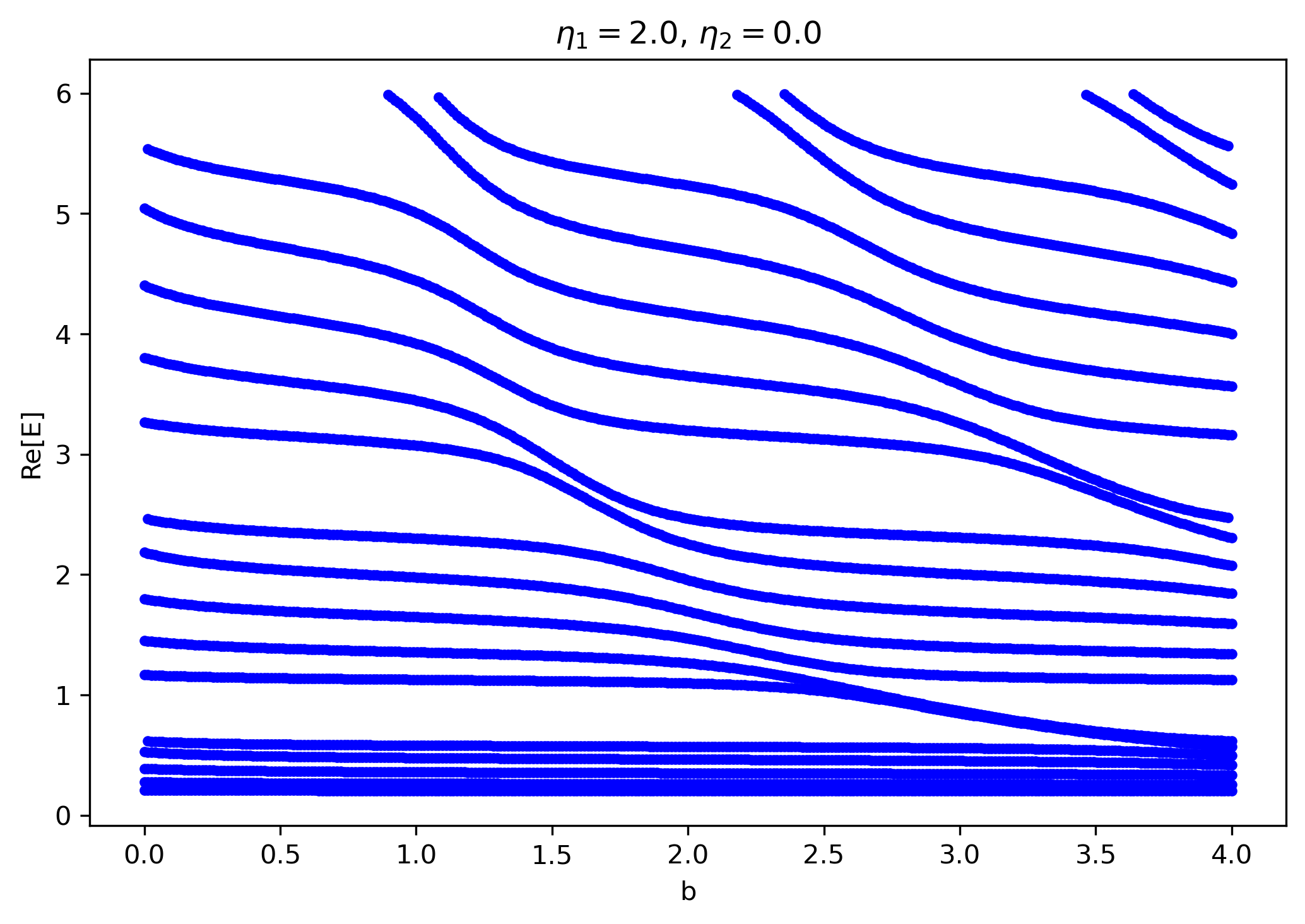}
\caption{$\eta_2=0$}
\end{subfigure}
\begin{subfigure}[b]{0.99\textwidth}
\includegraphics[width=0.995\textwidth]{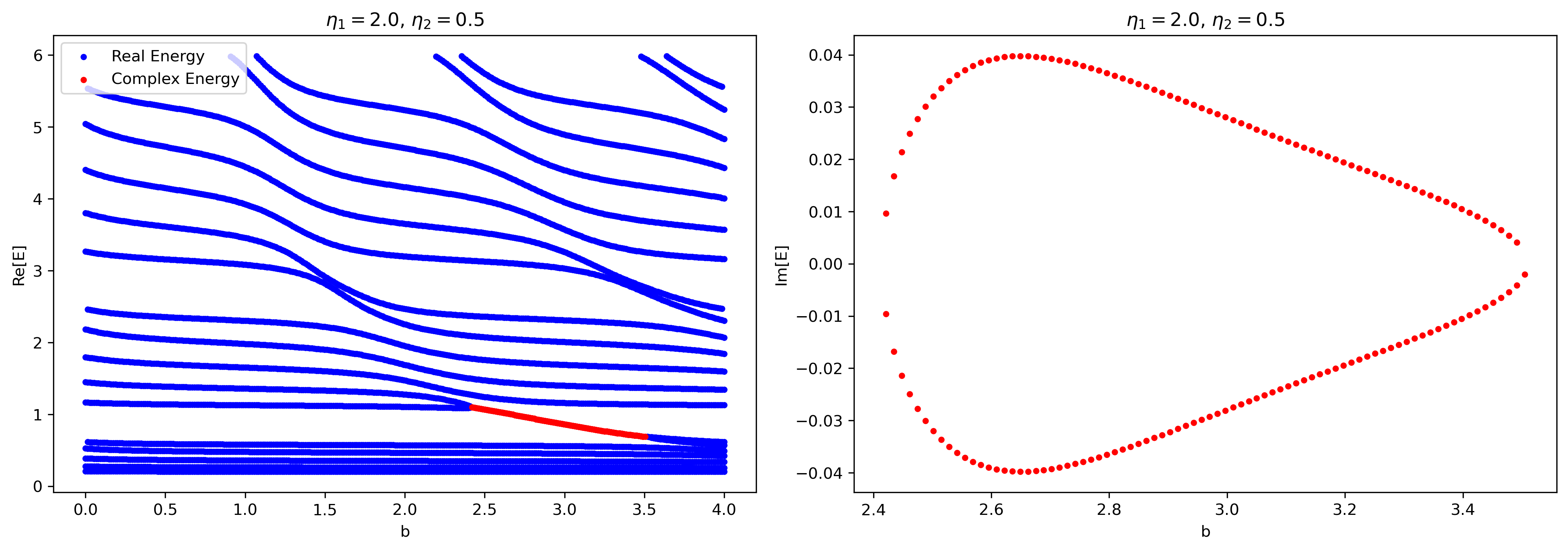}
\caption{$\eta_2=0.5$}
\end{subfigure}
 \begin{subfigure}[b]{0.99\textwidth}
\includegraphics[width=0.995\textwidth]{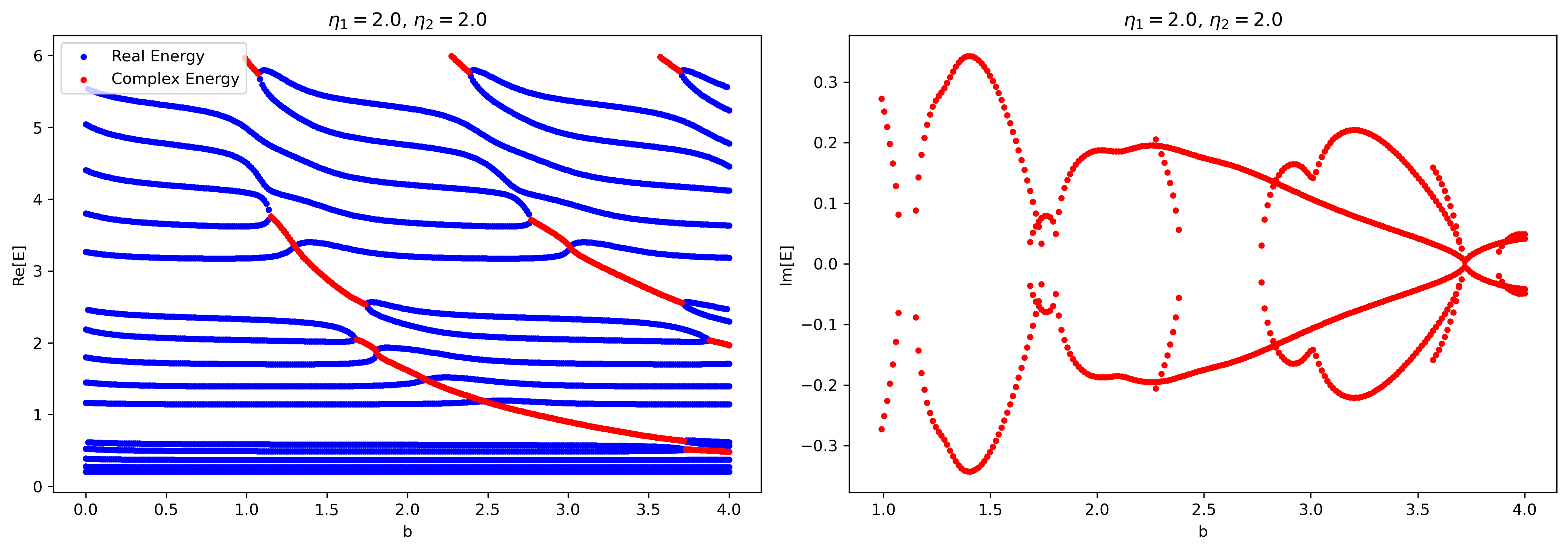}
\caption{$\eta_2=2$}
\end{subfigure}
\caption{The real and imaginary parts of energy spectrum as a function of the $b$ with hard wall boundary conditions placed at $x_0=0$ and $x_{N+1} =L$. The real energy solutions are plotted in blue, and the complex energy solutions are plotted in red. The four $V_0=1$ potentials are placed in between $z_1=\frac{\eta_1-i\eta_2}{2k}$ and $z_N=\frac{\eta_1+i\eta_2}{2k}$, the spacing between two potentials is $a= 4$, $m=4$ and the  $\eta_1 = 2.0$, $\eta_2=0$ (upper panel) and $\eta_1= 2.0$, $\eta_2=2.0$ (lower panel).}
\label{Fig.6}
\end{figure*}

\begin{figure*}
 \begin{subfigure}[b]{0.99\textwidth}
\includegraphics[width=0.495\textwidth]{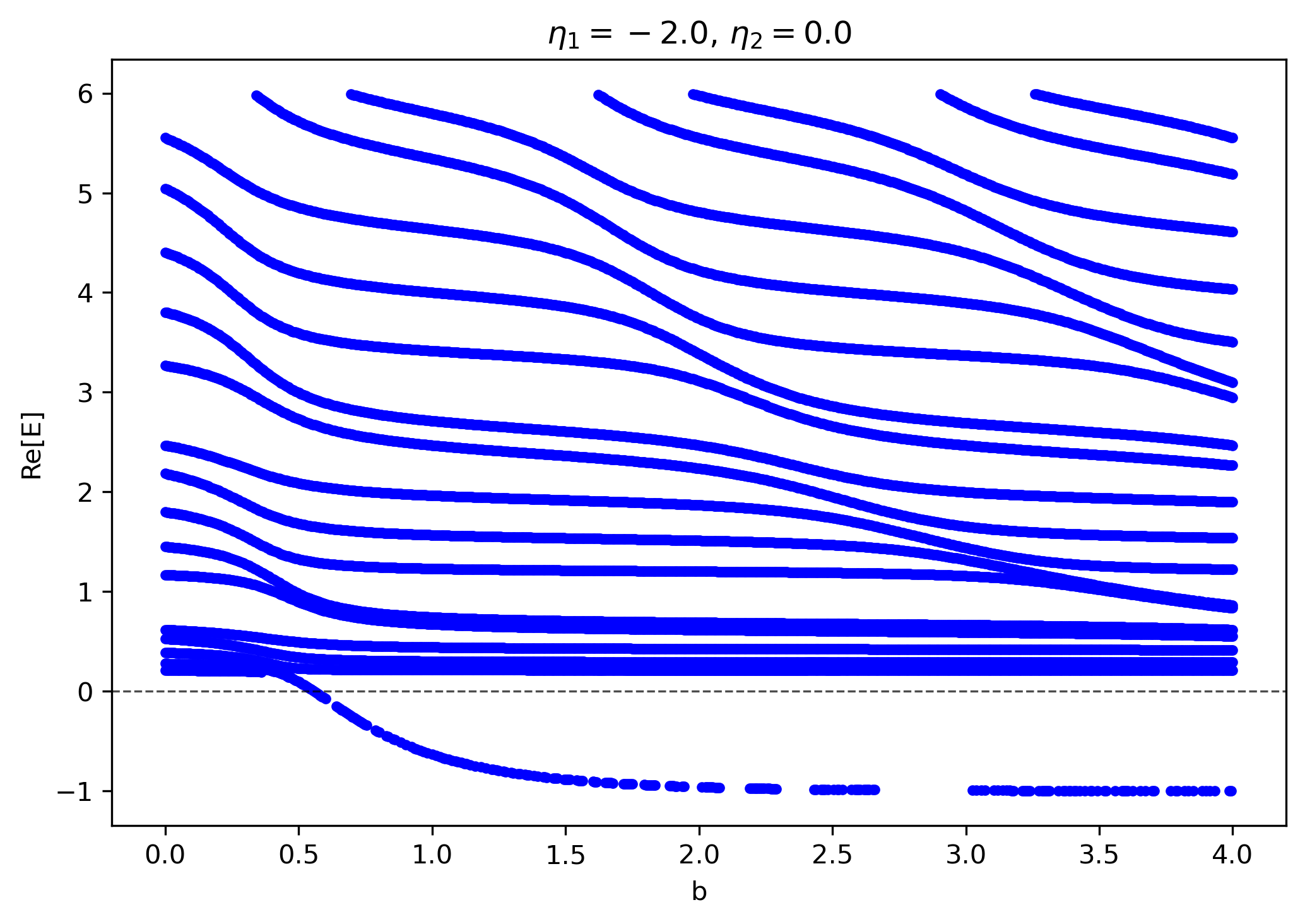}
\caption{$\eta_2=0$}
\end{subfigure}
\begin{subfigure}[b]{0.99\textwidth}
\includegraphics[width=0.995\textwidth]{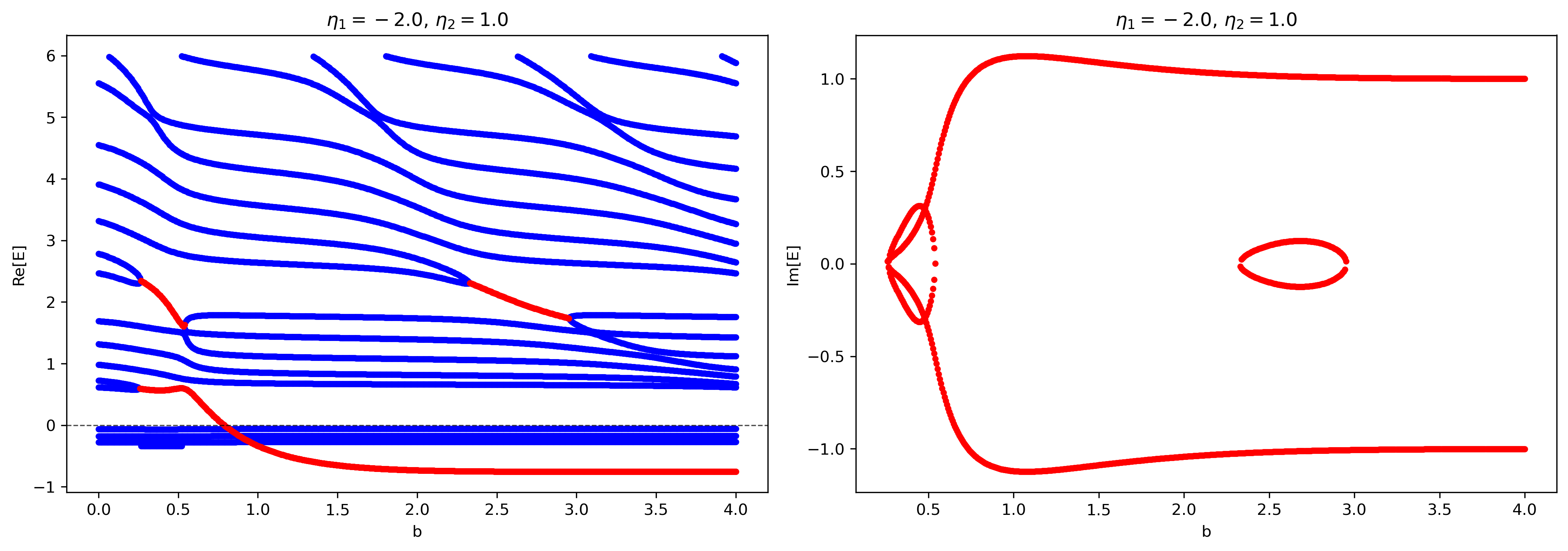}
\caption{$\eta_2=0.5$}
\end{subfigure}
 \begin{subfigure}[b]{0.99\textwidth}
\includegraphics[width=0.995\textwidth]{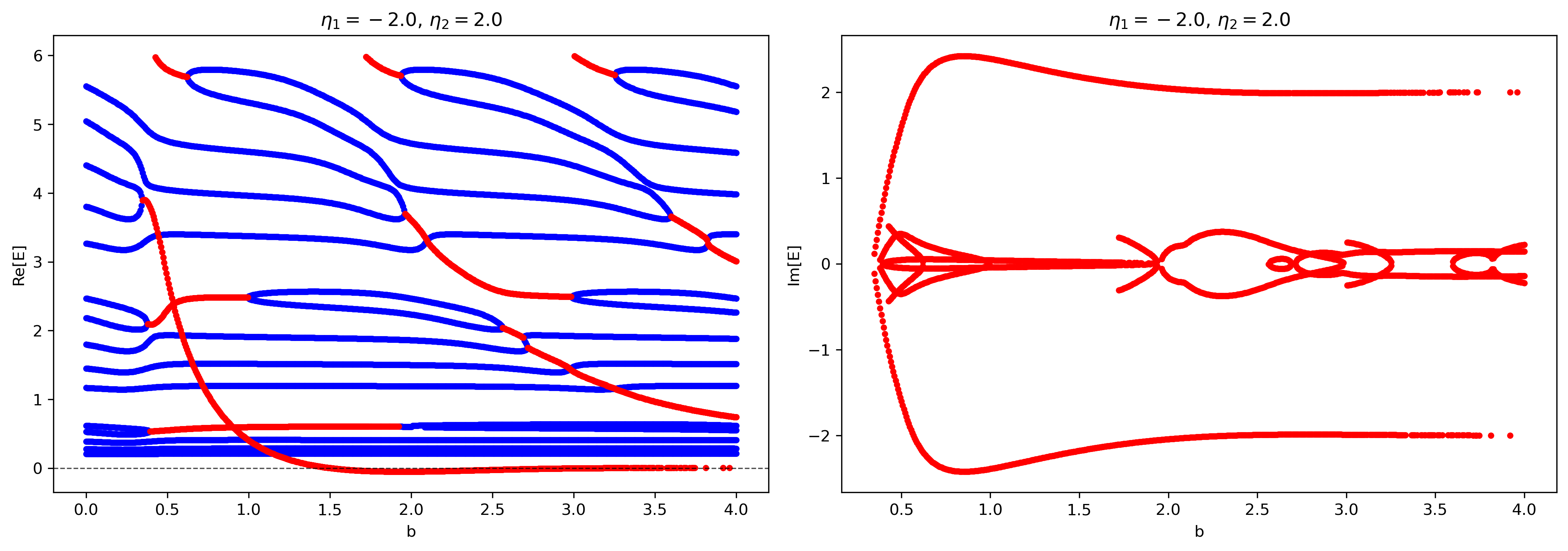}
\caption{$\eta_2=2$}
\end{subfigure}
\caption{Same as in Fig. 6, but for a system of negative barriers with $\eta_1=-2$ and various values of $\eta_2$.}
\label{Fig.7}
\end{figure*}

In Fig. \ref{Fig.5} we observe the emergence of new regions with increasing $\eta_1$, where $T_N$ becomes greater than $1$. The physical reason for this difference is connected with the fact that the system enters the region where some eigenvalues become complex, and complex conjugate pairs (see Refs. \cite{ahmed1,Gas2025} for more details.) In the
mathematical sense, this is explained by the more complex expression for the ranges (see the two-sided inequality  (\ref{t13})), which now also depends on $\eta_1$.

Fig.\ref{Fig4} shows the dependence of the transmission coefficients $T_N$, Eq.(\ref{ty}), the reflection coefficients on the left $R_L$ and on the right $R_R$, Eq. (\ref{r_LR}), versus $k$ at $\eta_1=0$. At first glance they are very different in detail, however, since the denominator of all three quantities is the same, they have anomalous behavior in the same energy range, mainly defined by the inequality (\ref{t13}). 

First of all, note that the asymmetric behavior of $R_{L/R}$ is essentially determined by the sign of the potential $\eta_2$ that the incident wave first meets. As a consequence the area covered by $R_L$ is always larger than the area covered by $R_R$. Consequently the flux from the left ($\eta_2>0$) is enhanced and $T+R_L>1$. The flux from the right ($\eta_2<0$) is suppressed and $T+R_R<1$. Usually, this type of non-standard behavior of scattering matrix elements is connected with a very specific spectral singularity of discrete energy spectrum (see, e.g., Ref. \cite{ahm1}).  These discrete energies can be real negative, real positive and 
complex conjugate pair(s) of eigenvalues. The technical aspects of determining the real positive energy values of critical energies associated with spectral singularities for the hybrid system are discussed in detail in subsection $C$ (see Fig.\ref{Fig2} for $k=1.39$ ). 

\section{Energy spectrum}

In the present section, we investigate the energy spectrum of the hybrid system described above (see Eq. (\ref{pot2})). We will close the system by imposing boundary conditions in the form of hard walls at points $x_0$ and $x_{N+1}$, adding two extra real delta potentials $V_0$ and $V_{N+1}$ . We will argue later on that when $V_0$ and $V_{N+1}$ tend to $\infty$ the properties of the system are obviously equivalent to the properties of a closed system. 
In fact, this model exhibits the properties of a closed system already when $\frac{V_0}{2k}\gg 1$ and $\frac{V_{N+1}}{2k}\gg 1$.
Remarkably, as we will see below, within the characteristic determinant approach quite different
physical behaviors of the hybrid system from open to closed is highly sensitive to
the initial conditions for the recurrence relations. Let us recall that the finite hybrid structure, the subject of this article,
consisting of a real potential region (passive region with $n$ $\delta$-potentials and with $V_0$ real strengths) surrounded by gain and loss regions on the left and right, respectively (see Eq. (\ref{pot2})). The gain and loss regions are presented by complex delta potentials $z_1=\frac{\eta_1+i\eta_2}{2k}$ and $z_N=z^*_1=\frac{\eta_1-i\eta_2}{2k}$, respectively. They are located at $x=x_1$ and $x=x_N$.  The hard wall boundary conditions are fixed at $x_0$ and $x_{N+1}$. Let us denote $a$ as the distance between two successive real potentials in the passive region, and $b$ as the distance between hard wall boundaries and the first $z_1$ and $z_N$ last complex potentials, that is $x_1-x_0 =x_{N+1}-x_N\equiv b.$ The parameter $b$, as we will see below, plays an important role in moving the boundary of the entire system and, in essence, changing the size of the system. It will also play a decisive role in shaping the energy spectrum within a finite-sized box.

The basic idea is to write down, in a similar way to that obtained in (\ref{e:d0}), the new recurrence relation for the characteristic determinant,  when $V_0$ and $V_{N+1}$ tend to $\infty$. Then, once the $\mathcal{D_{N}}$ determinant is found, one can straightforwardly investigate in details its zeros. Further manipulations are
completely analogous to those outlined in Section II for the open system. Thus, as a result for the energy spectrum we arrive at (in the text below we use $z_m$ instead of $V_m$, implying that the amplitude may also be complex).
\begin{equation}
\bigg(i\frac{z_m}{2k}(1-\lambda^2_{m+1})+\lambda^2_{m+1}\bigg)\mathcal{D}_m-{\lambda^2_{m+1}}\mathcal{D}_{m-1}\equiv 0, \label{km}
\end{equation}

with the following recurrence relation for determinant $\mathcal{D}_m$
\begin{equation}
\mathcal{D}_{m}=\mathcal{A}_m\mathcal{D}_{m-1}-\mathcal{B}_m\mathcal{D}_{m-2},\label{rec}
\end{equation}
where $\mathcal{A}_m=1+i\frac{z_m}{2k}(1-\lambda^2_m)+\frac{z_m}{z_{m-1}}\lambda^2_m$ and $\mathcal{B}_m=\frac{z_m}{z_{m-1}}\lambda^2_m$, $\lambda_m=\exp{ik(x_m-x_{m-1})}$ with 
initial conditions:
$\mathcal{D}_0=1$, $\mathcal{D}_1=1+i\frac{z_1}{2k}(1-\lambda^2_1)$.

It is worth noting that by substituting $\mathcal{D}_0$ and $\mathcal{D}_1$
into equation (\ref{km}), one can directly obtain the quantization condition
for the single $\delta$ potential in the infinite well \cite{1d,2d}. By calculating $\mathcal{D}_2$ from the recurrence relation (\ref{rec}) and substituting $\mathcal{D}_2$ and $\mathcal{D}_1$ into Eq. (\ref{km}), we obtain the bound state solutions of $\mathcal{PT}$ -symmetric diatomic molecule (see Eq. (\ref{km13}) below)).  In particular, in the case $\eta_2=0$ we obtain the result discussed, for example, in the Ref. \cite{tommy}). By adding a new delta potential (real or complex) and solving Eq. (\ref{km})  (numerically or analytically), one can find the quantization condition of the elementary excitations (electrons, photons, etc) in a system with a finite number
of point barriers of arbitrary height, located in arbitrary
positions (see Eq. (\ref{pot2})). 

Let us emphasize once again that idealized closed systems, described exclusively by the Schr{\"o}dinger equation, and open systems, described by the non-Hermitian Hamiltonians, differ from each other only in the initial conditions of the characteristic determinant, that is, $D_1=1+i\frac{z_1}{2k}$ and $\mathcal{D}_1=1+i\frac{z_1}{2k}(1-\lambda^2_1)$ respectively.  
\subsection{Analytical approach}
To further investigate the energy spectrum of the hybrid system (\ref{km}), we can similarly to the description above, using the recurrence relations (\ref{rec}) for the characteristic determinant $\mathcal{D}_m$, applied to both ends, rewrite it as follows ($m=N-2$) (see Refs. \cite{GAM,Gas2025}):
Before doing so, it is instructive to calculate $\mathcal{D}_m$
in case of closed periodic system. 
In order to arrive at the desired expression for $\mathcal{D}_m$, we  
closely follow Refs. \cite{GA88,GAM}. 
Assuming that the quantities $\mathcal{A}_m$ and $\mathcal{B}_m$ (see (\ref{rec})) do not depend on the index $m$, the recurrence relation for $\mathcal{D}_m$ can be reduced to a quadratic equation with constant coefficients. After some algebraic 
manipulations, we obtain the following expression: 
\begin{equation}
\mathcal{D}_m
=e^{ikm}\bigg[\cos m\beta a+
\bigg(\frac{V_0}{2k}-i\bigg)\frac{\sin ka\sin {m\beta a}}{\sin \beta a}\bigg] \label{Dm}.
\end{equation}
The above expression for $\mathcal{D}_m$ shows some similarity with $D_m$ (see (\ref{Mn})), since it also describes a periodic structure. However, it should be noted that the structure of $\mathcal{D}_m$ in the case of a closed system, despite the formal similarity, is very different from the structure of $D_m$ calculated for an open system. The physical reason for this difference lies in the type of system under consideration. 

To demonstrate this explicitly, consider a closed system with a point source on the left hard wall, i.e., at the point $x = x_0$. The transmission coefficient $\mathcal{T}_m$ at the point $x_{N+1}$, where the right boundary is located, can be expressed similarly to Eq. (\ref{tz}) as follows:
\begin{equation}
\mathcal{T}_m=\frac{1}{|\mathcal{D}_m|^2}=\frac{1}{1+\frac{V_0}{k}\sin{ka}\frac{\sin {m\beta a}}{\sin \beta a}\frac{\sin {(m+1)\beta a}}{\sin \beta a}}\label{Tm1}.
\end{equation}

Comparing the two formulas, we see
that in Eq. (\ref{Tm1}) there are 
two important energy scales, for which an incident wave is totally transmitted, i.e., $\mathcal{T}_m
=1$. The first case, which is typical for an open system (see Eq. (\ref{tz})) , occurs when $\frac{\sin(m\beta a)}{\sin(\beta a)}=0$. This corresponds to constructive interference between paths reflected from $m$ $\delta$ potentials located in different places. We get, therefore ($n$ is the order of the resonance mode):
\begin{equation*}
\beta a =\frac{\pi n}{m}, \qquad {n=1,2,... (m-1)}.
\end{equation*}

In the second case, which is clearly related to the presence of hard walls at $x=x_0$ and $x=x_{N+1}$, the desired extra resonance spectral peaks appear under the condition $\frac{\sin(m+1)\beta a}{\sin(\beta a)}=0$, which leads to 
\begin{equation*}
\beta a =\frac{\pi n}{m+1}, \qquad {n=1,2,... m}.
\end{equation*}

 Clearly, one then expects that, in the limit of a large number of $\delta$ potentials ($m\gg 1$), both spectral peaks defined above coincide.
This means that in this limit the boundary effects can be neglected and one can simply use $T_m$ defined by Eq. (\ref{tz}). 
However, at small values of $m$ the situation is somewhat more complicated: we observe two distinct resonance peaks.
The situation is analogous to resonant tunneling through a multilayer structure,  superlattices or structures that combine two Fabry-Perot interferometers where several spectral peaks can be observed simultaneously (see, e.g., \cite{13,14}). 

To proceed further and find the analytical expression for the energy spectrum of the 
hybrid system, 
let us insert the explicit expression of $\mathcal{D}_m$ (\ref{Dm}) into the recurrence relation (\ref{rec}). By solving relation (\ref{rec}), one obtains after some tedious algebra the following 
expression for the energy spectrum of the hybrid
closed system with hard wall boundary conditions:
\begin{multline}
\frac{1}{\sin{\beta a}}\bigg\{
\frac{\eta^2_1+\eta^2_2}{2k^2}\sin{ka}\sin^2{kb}\sin{(m+1)\beta a}
+\frac{\eta_1}{k}\sin{kb}\bigg[{\sin{(m+1)\beta a}}\sin{k(a+b)}-
{\sin{m\beta a}}\sin{kb}
\bigg]\\
-\sin{k(a+2b)}\frac{\sin{(m-1)\beta a}}{2}+{\sin{{m}\beta a}}\frac{\sin{k(a+b)}}{\sin{ka}}\bigg(\cos{\beta a}\sin{k(a+b)-\sin{kb}}\bigg)\bigg\}\equiv 0
, \label{km12} 
\end{multline}
Equation (\ref{km12}) is our main general result for the energy spectrum of $\mathcal{P}\mathcal{T}$-symmetric hybrid finite systems.
 It holds for both real and complex potentials, extending several well-known results in the literature, and helps to get even more insight into the mathematical structures of the eigenvalues. 
This was clearly demonstrated in recent Ref. \cite{Gas2025}, using the relatively simple finite bipartite Kroning-Penney model,
with complex potentials of periodic sets of two $\delta$ potentials in the unit cell.

It is easy to show that in case $\eta_2=0$, $\eta_1=V_0$ and $b=a$ Eq. (\ref{km12})  
simply reduces to ${\sin{ka}}\frac{\sin{(m+3)\beta a}}{\sin{\beta a}}=0$.
The latter is the quantization condition for $m+2$ equally spaced scatterers of the same height $V_0$ in an infinite potential well \cite{2d,Gas2025}. 
In the case $m=0$, Eq. (\ref{km12}), after simple
algebraic manipulations, can be rewritten in the form:
\begin{equation}
\frac{\eta^2_1+\eta^2_2}{2k^2}\sin{ka}\sin^2{kb}+\frac{\eta_1}{k}\sin{kb}\sin{k(a+b)}+\frac{\sin{k(a+2b)}}{2}
=0. \label{km13}
\end{equation}
It can be shown that the above expression for the energy spectrum can be obtained directly from (\ref{rec}),
after substituting $\mathcal{D}_2$ and $\mathcal{D}_1$ and solving for bound states. Note that Eq. (\ref{km13}) was analyzed in detail analytically and numerically in \cite{Gas2025}.

Now we turn to analyzing the energy spectrum expression Eq. (\ref{km12}) for
different $\eta_1$ and $\eta_2$, based on relation in Eq. (\ref{spectra}), to see the
dynamical evolution of the band structure depending on the
imaginary portion $\eta_2$. Depending on the sign of $\eta_1$ the
following two situations are to be distinguished.
\subsubsection{Passive region contains $m$ positive $\delta$ potentials}
The spectrum of the hybrid system as a function of the parameter $b$ is presented in Fig. 6.
First of all, we note that the number of levels $N$ in each allowed zone—regardless of the specific value of $\eta_2$—differs between the cases $b=0$ and $b=a$ due to a change in the system size. In the case $\eta_2=0$, the spectrum consists of
real eigenenergies,
with the value of $N$ varies, as $b$ increases, within the range from a minimum value of $(m+1)$ to a maximum value of $(m+3)$.
Now, let us first look at what happens to the band structure
when $\eta_2\ne 0$. It
is clear that the small values of $\eta_2$ ($\eta_2\ll \eta_1)$ do not change
the general shape of the band structure, but they do change
the values of the energy levels. As a result, one obtains the
usual band structure for the complex potential as one would
expect for a real potential (Fig.6 (a)). However, as seen
from Fig. 6(b) ($\eta_2 = 0.5$), some states
are already missing with increasing strength of the $\eta_2$ (see the red
line in Fig. 6(b)).
This
applies mainly to the degenerate states that have separated from the upper band and move to lower band through the
forbidden gap, accurately tracking the bulk bands trajectories. As $\eta_2$ increases the system enters the region where more eigenvalues become complex, and complex conjugate pairs (see
Fig. 6(c)). The latter means that the
number of real eigenenergies has decreased. With the further increasing of $\eta_2$ the number of complex levels
increases. As a result, more states begin to move out from
the given region, and further modification of band structures
occurs.
\subsubsection{Passive region contains $m$ negative $\delta$ potentials}
A completely different situation is observed in the case of negative values of $\eta_1$, since for certain values of $\eta_2$ (see Figs. 7(b) and (c)), the topological states characteristic of $\eta_2 = 0$ case disappear. These states detach from the upper band, move to the lower band through the forbidden gap,
and remain invariant under continuous deformations (see Fig. 7(a)). The effective mass of these states is very large due to the central position and, generally, they are insensitive to local perturbations, or in other words, any adiabatic deformation that respects certain symmetries of the system will not affect the existence of the symmetry-protected edge states. 
Now, let us first look at what happens to the band structure
when $\eta_2 \ne  0$. It is clear that the small values of $\eta_2$ ($\eta_2\ll\eta_1$) do not change
the general shape of the band structure, but they do change
the values of the energy levels. As a result, we obtain a band structure very similar to that shown in Fig. 7(a). However, as seen
from Fig. 7(b) and (c), with increasing strength of the $\eta_2$, some states
are already missing (red lines). This
applies mainly to the degenerate edge states, discussed above.  
The dynamics of the subsequent band structure as $\eta_2$ increases are quite similar to those we observed in the case of positive delta potentials.

In conclusion, we present one further illustrative example in which the advantage of the approach based on the recurrence relation (\ref{rec}) is successfully leveraged to derive a compact analytical expression for the quantization condition that determines the energy spectrum of a model corresponding to the placement of a rigid lattice within a finite-sized box (see, for example, Refs.\cite{resh,he}). The model can be parameterized by the lattice's position as follows: $x_l=-\frac{L}{2}+(l+\frac{\Delta-1}{2})\frac{L}{N}$ and $x_l \in [-\frac{L}{2},+\frac{L}{2}].$ $L$ represents the length of the infinite square well, $N$ is the number of delta potentials ($l=1,2 \ldots, N$) and $a=\frac{L}{N}$ is the lattice constant. The parameter $\Delta$ varying within the interval $[-1,+1]$ is a shift in the barrier positions with respect to the walls of the box. 
The energy spectrum of the equidistant scatterers of equal heights $V_0$
and with the position $x_l$ of the lattice defined above, can be presented in a
 closed-form expression in terms of the number of cells $N$ and parameters $b_1=\frac{\Delta+1}{2}\frac{L}{N}$
 and  $b_N=\frac{1-\Delta}{2}\frac{L}{N}$. The last ones, as we will see below,play an important role in moving a rigid lattice from the  
leftmost barrier at $\Delta=-1$ to the rightmost
barrier when $\Delta=+1$. 
By solving the recurrence relation (\ref{rec}) for equidistant scatterers with real potentials of equal height $V_0$ and coordinates $x_l$ distributed in accordance with the model described above, we obtain the following compact expression for the quantization condition ($\Delta a\equiv b_1-b_N$):
\begin{equation}
\bigg[\frac{V_0}{2k}\bigg(\cos{ka}-\cos{\Delta ka}\bigg)-\sin{ka}
\bigg]\frac{\sin{(N+1)\beta a}}{\sin{\beta a}}=0, \label{new1}
\end{equation}
A characteristic and important feature of this model (as reported in Refs. \cite{resh,he} based on numerical calculations) is that, within each zone, the lower ($N-1$) levels are independent of the shift parameter $\Delta$ and the distance between them increases with increasing $N$. As for the $N$-th level is situated in the adjacent band gap, corresponds to a topological edge state, and exhibits sensitivity to $\Delta$. The physical nature of this feature is readily apparent from the analytical expression (\ref{new1}), that fully explains the obtained numerical results of Refs. \cite{resh,he}. Indeed, the second factor, containing the number of delta potentials $N$, does not depend on the shift parameter $\Delta$. Unlike the second factor, the first contains a shift parameter $\Delta$ and describes the behavior of an $Nth$-level situated in the subsequent gap. This level oscillates and corresponds to a topological edge state. As $k$ increases, that is, when the Nth level situated within higher forbidden gaps, the number of oscillations increases. 

The numerical analysis presented in Refs. \cite{resh,he} shows that, for example, the first topological edge state lies in the range of approximately $5$ to $7$ (in appropriate units $\hbar^2/m=1$). These values coincide with the zeros of the first factor of Eq. (\ref{new1}) $\frac{V_0}{2k}(\cos{ka}-\cos{\Delta ka})-\sin{ka}
=0,$
at $\Delta=\pm 1$ ($k=\pi$) and $\Delta=0$ ($k\approx 3.7$).
The latter constitutes a solution to a transcendental equation $\frac{V_0}{2k}\sin{\frac{ka}{2}}+\cos{\frac{ka}{2}}=0$ at $V_0=.4$ and $a=1$.

Note, that the formula (\ref{new1}) reduces to the $\sin{ka}\frac {\sin{(N+1)\beta a}}{\sin{\beta a}}=0$ expression, as one would expect, when $\Delta =\pm1$.

\section{SUMMARY AND OUTLOOK}

By calculating the zeros of the characteristic determinant (or the poles of the Green's function), we obtained a closed-form expression for the energy spectrum of hybrid $\mathcal{PT}$-symmetric systems, consisting of a region with a real potential (a passive region) bounded on the left and right by a pair of complex-conjugate $\delta$-function potentials. It has been shown that under certain conditions and a specific ratio between the
real and imaginary parts of the complex potentials, it is possible to find analytical expressions for the spectral
singularities at which the scattering matrix elements of the hybrid structure tend to infinity at a specific real
energy.
Within the framework of the same approach, we present a compact analytical expression for the quantization condition that determines the energy spectrum of a model corresponding to the placement of a rigid
lattice within a finite-sized box. To the best of our knowledge, until now, this model has been accessible for study solely numerically.

\section{ACKNOWLEDGMENTS}
We would
like to thank UPCT for partial financial support through
“Maria Zambrano ayudas para la recualificación del sistema universitario español 2021–2023” financed by Spanish
Ministry of Universities with funds “Next Generation” of
EU.

\bibliography{ALL-REF.bib}

\end{document}